\documentclass[12pt,a4paper]{article}
\usepackage{graphics,graphicx}
\usepackage{amssymb,epsfig,amsmath,euscript,array}
\usepackage{cite}

\begin{document}

\begin{flushright}
hep-th/0609011\\
QMUL-PH-06-10
\end{flushright}

\vspace{20pt}

\begin{center}

{\Large \bf Quantum MHV Diagrams}
%
\vspace{33pt}

{\bf {\mbox{Andreas  Brandhuber and Gabriele  Travaglini}}}%
\footnote{{\sffamily \{\tt a.brandhuber, g.travaglini\}@qmul.ac.uk }}\\
\vspace{12pt}
{\em Department of Physics\\
Queen Mary, University of
London\\
Mile End Road, London, E1 4NS\\
United Kingdom}
\vspace{40pt}

{\bf Abstract}

\end{center}


\noindent
Over the past two  years, the use of on-shell techniques
has deepened our understanding of the $S$-matrix of gauge theories
and led to the calculation of many new scattering amplitudes.
In these notes we review a particular on-shell method developed recently,
the quantum MHV diagrams, and discuss applications to one-loop amplitudes.
Furthermore, we briefly discuss the application of $D$-dimensional generalised unitarity
to the calculation of scattering amplitudes in non-supersymmetric
Yang-Mills.

\vspace{0.5cm}

\setcounter{page}{0}
\thispagestyle{empty}
\newpage


\section{Introduction}

Recently, remarkable progress  has been made in 
understanding the structure of the $S$-matrix of 
four-dimensional gauge theories. 
This progress was prompted by Witten's proposal  \cite{witten}  of a new duality
between $\mathcal{N} \!\! = \!\! 4$ supersymmetric Yang-Mills 
(SYM) and the topological open string B-model with target space the Calabi-Yau 
supermanifold  $\mathrm{\bf CP}^{3|4}$, a supersymmetric version of Penrose's
twistor space. In contrast to usual dualities, this novel
duality relates two weakly-coupled theories, and as such it can in principle
be tested by explicit computations on both sides.

One of the striking results of Witten's analysis is 
the understanding of the remarkable simplicity of scattering amplitudes 
in Yang-Mills and  gravity -- simplicity which is 
completely obscure in computations using textbook techniques -- 
in terms of the geometry of twistor space.
More precisely, Witten \cite{witten} observed that tree-level scattering amplitudes, 
when Fourier transformed to twistor space, localise on algebraic curves
in twistor space.
For the simplest case of the maximally helicity violating (MHV) amplitude,
the curve is just a (complex) line.

The simple geometrical structure in twistor space of the
amplitudes was also the root of further important developments of
new efficient tools to calculate amplitudes. In \cite{csw1},
Cachazo, Svr\v{c}ek  and Witten (CSW) proposed a novel
perturbative expansion for  amplitudes in YM, where the
MHV amplitudes are lifted to vertices, and joined by scalar
propagators in order to  form amplitudes with an increasing number
of negative helicity gluons. 
At tree level, numerous successful applications of the MHV diagram 
method have been carried out so far 
\cite{Zhu}--\!\!\cite{Ozeren:2005mp}.    
An elegant proof of the method 
at tree level was presented in \cite{bcfw} based on 
the analyticity properties of the scattering amplitudes. 
Later, this method was shown to be
applicable to one-loop amplitudes in supersymmetric
\cite{bst,bbst1,qr} and non-supersymmetric theories \cite{bbst2},
where a new infinite series of amplitudes in pure YM was
calculated. In a different development, the new twistor ideas were
merged with earlier applications of unitarity
\cite{bdk1,bdk2} and generalised unitarity \cite{a}, and led to
highly efficient techniques to calculate one-loop amplitudes in
$\mathcal{N}=4$ SYM \cite{bcf-gen,bdk-december} and in
$\mathcal{N}=1$ SYM \cite{BBDP,BBCF}.

There are many important reasons for the interest in new, more
powerful techniques to calculate scattering amplitudes. Besides
improving our theoretical understanding of gauge theories at the
perturbative level the most important reason is the need for
higher precision in our theoretical predictions. The advent of the
Large Hadron Collider  requires the knowledge of perturbative
QCD backgrounds at unprecedented precision to distinguish ``old"
physics from the sought for ``new" physics, and there exist long
wishlists of processes that are yet to be computed.
Traditional methods using Feynman rules are rather inefficient
since they hide the simplicity of scattering amplitudes,
intermediate expressions tend to be larger than the final formulae
and one has to face the problem of the factorial growth of the
number of diagrams which hampers the use of brute force methods.
Therefore, techniques that directly lead to the simple final
answers are desirable. 
In the following we want describe some of the
novel ``twistor string inspired" techniques,  
focusing on methods relevant for one-loop amplitudes.

\section{Colour Decomposition and Spinor Helicity Formalism}

Here we describe two ingredients  that are essential in order to 
make manifest the simplicity of scattering amplitudes: 
the colour decomposition, and the spinor helicity formalism. 
We will later see how new twistor-inspired techniques merge fruitfully 
with these tools. 

At tree level, Yang-Mills interactions are planar, hence an
amplitude can be written as a sum over single-trace structures
times partial or colour-stripped amplitudes,
\begin{eqnarray}
{\mathcal A}_{n}^{\rm tree} \left( \{
p_{i},\epsilon_{i},a_{i}\}\right)&
=&
\sum_{\sigma \epsilon
S_{n}/{\bf Z}_{n}} \mathrm{Tr}\left( T^{a_{\sigma(1)}} \ldots
T^{a_{\sigma(n)}} \right) \\ \nonumber &&
A_{n} \left( \sigma(p_{1},\epsilon_{1}), \ldots,
\sigma(p_{n},\epsilon_{n}) \right) \, .
\end{eqnarray}
Partial amplitudes do not carry any colour structure. They
receive contributions only from Feynman diagrams with a fixed
cyclic ordering of the external lines and have a simpler analytic
structure than the full amplitude.
At loop level, also multi-trace structures appear, which are
subleading in the $1/N$ expansion. However, for the one-loop gluon
scattering amplitudes there exist a simple (linear) relation
between the planar and non-planar terms and, therefore, we have to
consider only planar ones.

The spinor helicity formalism, is largely responsible for the
existence of very compact formulas of tree and loop amplitudes in
massless theories. In four dimensions the Lorentz group is ${\bf
SL}(2,{\bf C})$ and a Lorentz vector is equivalent to a bi-spinor,
i.e.~the four-momentum can be written as a $2 \times 2$  matrix
$p_{a\dot{a}}= p_{\mu}\sigma^{\mu}_{a\dot{a}} $, 
$a, \dot{a} =1,2$, 
where $a$ and $\dot{a}$ are left and right handed spinor indices,
respectively. If $p_{\mu}$ describes the momentum of a massless, on-shell particle,
then $p^{2}=\mathrm{det} p_{a\dot{a}}=0$ and the matrix
$p_{a\dot{a}}$  can be written as a
product of a left and a right-handed spinor
\begin{equation}
p_{a\dot{a}}= \lambda_{a} \widetilde{\lambda}_{\dot{a}}\, .
\end{equation}
In real Minkowski space the left and right-handed spinors are
related by complex conjugation
$\widetilde{\lambda}=\pm \overline{\lambda}$. It is useful to
introduce the following Lorentz invariant brackets: $\langle i j
\rangle = \epsilon^{ab} \lambda^{i}_{a} \lambda^{j}_{b}$ and 
$[ j i  ] = \epsilon^{\dot{a} \dot{b}} \widetilde{\lambda}^{i}_{\dot{a}} 
\widetilde{\lambda}^{j}_{\dot{b}}$, in terms of which dot products of
on-shell four-momenta can be rewritten as $2 k_{i} \cdot k_{j} =
\langle i j \rangle [ji]$.

The simplest non-vanishing tree-level gluon
scattering amplitudes have exactly two negative helicity
gluons, denoted by $i$ and $j$, and otherwise only positive
helicity gluons. Since amplitudes with zero or one negative
helicity gluon vanish, these amplitudes are called maximally
helicity violating (MHV). Up to a trivial momentum conservation
factor they take the following form
\begin{equation}\label{mhv}
A_{\mathrm{MHV}}(\lambda_{i})=i g^{n-2} \frac{\langle i j
\rangle}{\langle 1 2 \rangle \ldots \langle (n-1) n \rangle\langle
n 1 \rangle} \, .
\end{equation}
The most remarkable fact about this formula, besides its
simplicity, is that it is holomorphic (no $\widetilde{\lambda}$s
appear), which has important consequences when the amplitude is
transformed to twistor space. 

\section{Twistor Space and MHV Diagrams}

In a nutshell, twistor space is obtained from the spinor variables
$(\lambda_{a}, \widetilde{\lambda}_{\dot{a}})$ by performing a 
``half Fourier transform" (FT), i.e.~a FT of one of the spinor
variables, say $\widetilde{\lambda}_{\dot{a}} \rightarrow
\mu_{\dot{a}}$. Twistor space (for complexified Minkowski space)
is a four-dimensional complex space $(\lambda_{a},\mu_{\dot{a}})$.
However, amplitudes turn out to be homogeneous functions of the
twistor variables and it is more natural to think about projective
twistor space $(\lambda_{a},\mu_{\dot{a}}) \sim (t \lambda_{a},t
\mu_{\dot{a}})$ with $t$ a non-zero complex number. 

In twistor
theory a central r\^{o}le is played by the incidence relation
\begin{equation}
\label{inc}
\mu_{\dot{a}}+x_{a\dot{a}} \lambda^{a}\, =\, 0
\ , 
\end{equation}
 which describes the
correspondence between Minkowski space and twistor space. In
particular if we fix a point in Minkowski space $x_{a\dot{a}}$
this leads to two complex equations that describe a line in
projective twistor space.
As mentioned earlier the MHV tree amplitudes are holomorphic,
except for the overall momentum conservation factor
$\delta^{4}(\sum \lambda_{i} \widetilde{\lambda}_{i})$, and
therefore their transformation to twistor space is particularly
simple:
\begin{eqnarray}
\label{holo}
&&\widetilde{A}_{\mathrm{MHV}}(\lambda,\mu) \sim
A_{\mathrm{MHV}}(\lambda) \int \prod_{i} d\widetilde{\lambda}_{i}
e^{i \mu_{i} \widetilde{\lambda}_{i} }
e^{i x \lambda_{i} \widetilde{\lambda}_{i}} \nonumber \\
&&\sim  A_{\mathrm{MHV}}(\lambda) \prod_{i} \delta(\mu_i+x
\lambda_{i}) \, .
\end{eqnarray}
Notice the appearance in \eqref{holo} of the incidence relation 
\eqref{inc}, 
which implies that MHV amplitudes are supported on a line in
twistor space. For more complicated amplitudes the localisation
properties are almost as simple, in particular an amplitude with
$Q$ negative helicity gluons localises on sets of $Q-1$
intersecting lines.
\begin{figure}[ht]
\begin{center}
\scalebox{0.5}{\includegraphics{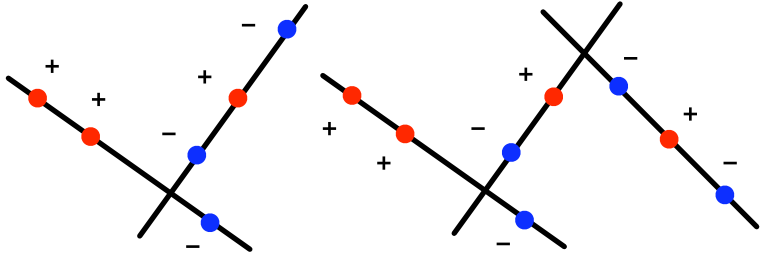}}
\caption{Twistor space localisation of tree amplitudes with $Q=3$
and $Q=4$} 
\label{nonmhv}
\end{center}
\end{figure}
These simple geometrical structures in twistor space of the
amplitudes led  Cachazo, Svr\v{c}ek  and Witten (CSW) to propose
\cite{csw1} a novel perturbative expansion for tree-level 
amplitudes in Yang-Mills using MHV amplitudes as effective
vertices. It is indeed natural to think of an MHV amplitude  as a local interaction, 
since the line in twistor space on which a MHV
amplitude localises corresponds to a  point in Minkowski
space via the incidence relation. 

In the MHV diagrammatic
method of \cite{csw1}, MHV vertices are connected by scalar
propagators $1/P^{2}$ and all diagrams with a fixed cyclic
ordering of external lines have to be summed. The crucial point is 
clearly to define an  off-shell continuation
of MHV amplitudes. 
CSW proposed to associate internal (off-shell) legs
with momentum $P$ the spinor  
$\lambda_{P a} := P_{a\dot{a}} \eta^{\dot{a}}$, 
where $\eta$ denotes an arbitrary reference spinor. 
We can assign a spinor to every internal
momentum and insert this spinor in the Parke-Taylor formula 
(\ref{mhv}) to define the off-shell MHV vertex. It can be shown
\cite{csw1} that after summing all diagrams the dependence on the
reference spinor drops out. Proofs of the equivalence of MHV diagrams 
and usual Feynman diagrams at tree level have been presented in 
\cite{bcfw,risager,mansfield,gr}. Interestingly, 
MHV rules for tree-level gravity have also been derived in 
\cite{swan}.

\section{From Trees to Loops}

The success of the MHV method at tree level brings up
the question if this can be extended to the quantum level, 
i.e.~to loop amplitudes. The original prognosis from twistor
string theory was negative because of the presence of 
unwanted conformal supergravity modes that spoil the duality 
with Yang-Mills at
loop level. Remarkably, we will find perfect agreement between 
loop MHV diagrams (such as in Figure 2) and the  
results obtained with more standard methods.


For simplicity let us focus on the simplest one-loop amplitudes,
the MHV one-loop amplitudes in $\mathcal{N}=4$ super Yang-Mills.
\begin{figure}[ht]
\begin{center}
\scalebox{1}{\includegraphics[height=9pc]{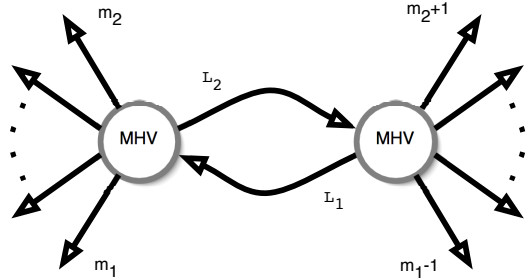}}
\caption{MHV diagrams for the MHV one-loop amplitudes in
$\mathcal{N}=4$ SYM} \label{fig:largenough1}
\end{center}
\end{figure}
These amplitudes were computed  in \cite{bdk1} using
the four-dimensional cut-constructibility approach, which utilises the fact that
one-loop amplitudes in supersymmetric theories can be
reconstructed from their discontinuities. The result is
surprisingly simple and can be expressed in terms of the so-called
{\it 2-mass easy box functions} $F^{\rm 2me}(s,t,P^{2},Q^{2})$ as
\begin{equation}
A^{\mathrm{1-loop}}_{\mathrm{MHV}}=A^{\mathrm{tree}}_{\mathrm{MHV}}
\times \sum_{p,q} F^{\rm 2me}(s,t,P^{2},Q^{2}) ~.
\end{equation}
\begin{figure}[ht]
\begin{center}
\scalebox{1}{\includegraphics[width=11pc]{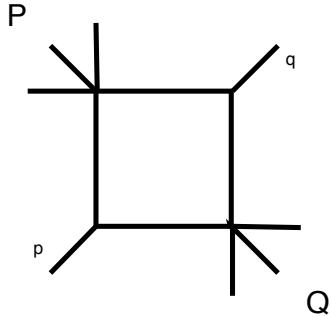}}
\caption{A graphical representation of the 2-mass easy box
function} \label{fig:largenenough2}
\end{center}
\end{figure}

It turns out that the MHV diagrammatic calculation \cite{bst}
agree perfectly with the result of \cite{bdk1}. A few remarks are
in order here:
\begin{itemize}
\item In the calculation it is crucial to decompose the loop
momentum $L$ in an on-shell part $l$ plus a part proportional to
the reference null-momentum $\eta$: $L = l + z \eta$.

\item This decomposition naturally leads to {\it dispersion
integrals} $\int dz/z (\cdots)$, which do not need subtractions. 

\item This calculation provides a check of the MHV method, 
e.g.~the proof of covariance ($\eta$-independence) is highly
non-trivial.

\item This calculation incorporates a large number of conventional
Feynman diagrams.
\end{itemize}

This approach readily applies to non-MHV amplitudes and theories
with less supersymmetry. In \cite{qr,bbst1} the method was applied
to the case of MHV one-loop amplitudes in $\mathcal{N}=1$ SYM and
complete agreement with a calculation using the
cut-constructibility approach \cite{bdk2} was found.
Finally, in \cite{bbst2} the  first new result from MHV diagrams
at one-loop was obtained: the cut-containing part of the MHV
one-loop amplitudes in pure Yang-Mills.

Having discussed these important checks of the  MHV method, in the next Section we will  
argue that generic one-loop scattering amplitudes 
can be equivalently computed with MHV diagrams.

\section{From Loops to Trees}

In order to prove that MHV diagrams produce the correct result for scattering
amplitudes at the quantum level, one has to

{\bf 1.} Prove the covariance of the result;

{\bf 2.} Prove that all physical singularities
of the scattering amplitudes (soft, collinear and multiparticle)
are correctly reproduced.

Indeed, if $A_{\rm MHV}^{(n)}$ is the result of the calculation of
an $n$-point  scattering amplitude based on MHV diagrams, and
$A_{\rm F}$ is the correct result (obtained using Feynman diagrams),
the difference $A_{\rm MHV}^{(n)} - A_{\rm F}^{(n)}$
must be a polynomial which, by dimensional analysis, has  dimension $4-n$.
But such a polynomial cannot exist except for $n=4$.
This case can be studied separately, and we have already discussed
the agreement with the known results for theories with
${\mathcal N}=1,2,4$ supersymmetry.

In this section we will will report on the calculation of generic one-loop
amplitudes with MHV diagrams \cite{bst2}. Firstly, we will
prove the covariance of the result.
This proof relies on two basic ingredients: 
the locality of the MHV vertices, and 
a beautiful result by Feynman known as the Feynman Tree Theorem
\cite{ofey}. This theorem relates the contribution of a loop
amplitude to those of amplitudes obtained by opening up the loop
in all possible ways; strikingly, this theorem allows one
to calculate loops from on-shell trees.
Our next goal will be to show that soft and collinear singularities
are correctly reproduced by MHV diagrams at one loop.
Thus, for a complete proof of the MHV method at one loop,
it would only remain to show the agreement of  multiparticle
singularities.
 
\subsection{The Feynman Tree Theorem and the Proof of Covariance}
We begin by briefly reviewing Feynman's Tree Theorem.
This result is based on the decomposition of the Feynman propagator
into a retarded (or advanced) propagator and a term which
has support on shell. For instance,
\begin{equation}
\label{ofey}
\Delta_F (P) \ = \ \Delta_R (P) \, + \, 2 \pi \delta (P^2-m^{2})
\theta (-P_0)
\ .
\end{equation}
Suppose we wish to calculate a certain one-loop diagram
$\mathcal{L}$, and let $\mathcal{L}_R$ be the quantity obtained from
$\mathcal{L}$ by replacing all Feynman
propagators by retarded propagators.%
\footnote{The same argument works for advanced propagators.}
Clearly $\mathcal{L}_R = 0$, as there are no closed timelike curves in
Minkowski space. This equation can fruitfully be used to work out 
the desired loop amplitude.
Indeed, by writing the retarded propagator as a sum of Feynman propagator plus
an on-shell supported term as dictated by (\ref{ofey}),
one finds that
\begin{equation}
\label{ftt2}
\mathcal{L} \ = \ \mathcal{L}_{\rm 1-cut}\, + \,
\mathcal{L}_{\rm 2-cut}\, + \, \mathcal{L}_{\rm 3-cut}\, +  \,
\mathcal{L}_{\rm 4-cut}
\ .
\end{equation}
Here $\mathcal{L}_{\rm {\it p}-cut}$ is the sum of all
the terms obtained by summing all possible diagrams obtained
by replacing $p$ propagators in the loop with delta functions.
Each delta function cuts open an internal loop leg,
and therefore a term with $p$ delta functions computes a
$p$-particle cut in a kinematical channel determined
by the cut propagators (whose momentum
is set on shell by the delta functions).

Feynman's Tree Theorem (\ref{ftt2}) states
that a one-loop diagram
can be expressed as a sum over
all possible cuts of the loop diagram.
The process of cutting puts internal lines
on shell; the remaining
phase space integrations have still to be performed,
but these are generically easier than the original loop integration.
Thus the Feynman Tree Theorem implies that one-loop
scattering amplitudes can be determined from on-shell data alone.
Interestingly, one can iterate this procedure and apply it to 
higher loop diagrams.

Now we apply the Feynman Tree Theorem to MHV diagrams.
This possibility is guaranteed by the local character
in Minkowski space of an MHV interaction vertex; thus
a closed loop MHV diagram where propagators are replaced 
by retarded or advanced propagators vanishes, and one arrives at
an equation identical to (\ref{ftt2}). Specifically, 
in \cite{bst2} we used Feynman's Tree Theorem in order to prove that one-loop
amplitudes calculated with MHV diagrams are covariant.
More precisely,
we were able to show that the sum of all possible $p$-cut MHV diagrams
is separately covariant. The remaining Lorentz-invariant phase space integrations are
also invariant, hence the full amplitude  -- given by summing over $p$-trees with
$p = 1, \ldots , 4$ as indicated by (\ref{ftt2}) --
is also  covariant.

A sketch of the proof of covariance for the case of one-loop
amplitudes with the MHV helicity configuration
is as follows. The one-loop MHV diagrams contributing to an
$n$-point MHV amplitude are presented in Figure 2.
\begin{figure}[ht]
\label{Figure3}
\begin{center}
\scalebox{0.75}{\includegraphics{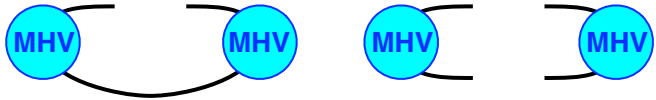}}
\end{center}
\caption{One-particle and two-particle MHV diagrams
contributing to the one-loop MHV scattering amplitude.
}
\end{figure}
In Figure 4 we show the one-particle and
two-particle cut diagrams which are generated in the
application of the Feynman Tree Theorem.%
\footnote{In our notation we only draw vertices and propagators
(or cut-propagators) connecting them. It will be understood that
we have to distribute the external gluons among the MHV vertices
in all possible ways compatible with cyclic ordering, and
the requirement that the two vertices must have
the helicity configuration of an  MHV amplitude.
Moreover we will have to sum over all possible
helicity assignments of the internal legs and, where required,
over all possible particle species which can run in the loop.
}
We start by focussing on one-particle cut diagrams.
These one-particle cut-diagrams are nothing but
tree-level diagrams, which are then integrated using a
Lorentz invariant phase space measure.
We now make the following important observation:
these tree (one-cut) diagrams would precisely sum to a
tree-level next-to-MHV (NMHV) amplitude
with $n+2$ external legs
(which would then be covariant as shown in \cite{csw1}),
{\it if} we also include the set of diagrams where
the two legs into which the cut propagator is broken
are allowed to be at the same MHV vertex.
Such diagrams are obviously never generated by cutting
a loop leg in MHV diagrams of the type depicted in Figure 2.
These ``missing'' diagrams are drawn in Figure 5.
MHV rules tell us, before any phase space integration is performed,
that the  sum of one-particle cut diagrams
of Figures 4 and 5
generates a NMHV amplitude with $n+2$ external legs.
Since the phase
space measure is Lorentz invariant, it follows that
the sum of one-particle cut diagrams,
including the missing diagrams, is covariant.

\begin{figure}[ht]
\label{Figure4}
\begin{center}
\scalebox{0.75}{\includegraphics{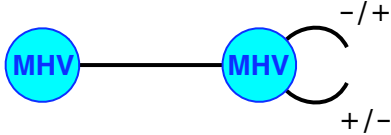}}
\end{center}
\caption{In this Figure we represent ``missing diagrams",
mentioned in the text.
}
\end{figure}

To complete the proof of covariance we have to justify 
the inclusion of these missing diagrams.
Here we will present an  explanation which 
relies on supersymmetry.%
\footnote{An alternative proof, not relying on supersymmetry, can be found 
in \cite{bst2}.}
The diagrams where two adjacent and opposite helicity
legs from the same MHV vertex are sewn together vanish
when summed over particle species in a supersymmetric theory.
Individual diagrams before summing over particle species diverge
because of the collinearity of the momenta of the two legs,
but the sum over particle species vanishes
even before integration.
So we discover that we could have actually
included these diagrams from the start,
since their contribution is zero.
Hence one-particle cut diagrams of
MHV one-loop amplitudes generate phase space integrals of
tree-level NMHV amplitudes, and are, therefore, covariant.

Next we look at two-particle cut diagrams.
These split the one-loop diagram of Figure 2
into two disconnected pieces (see the second diagram in Figure 4).
These are two MHV amplitudes, because the two internal legs
are put on shell by the Feynman cuts. Therefore,
no $\eta$-dependence is produced by these two-particle cut
diagrams.

Summarising, we have shown that Feynman one-particle and
two-particle cut diagrams are separately covariant. 
By Feynman's Tree Theorem (\ref{ftt2}) 
we conclude that the physical one-loop MHV amplitude is covariant too.

The main lines of the proof of covariance we have discussed 
in the simple example of an MHV amplitude are easily 
generalised more complicated amplitudes. In particular, 
we notice that: 

{\bf 1.} In a one-loop MHV diagram with $v$ vertices
and $n$ external particles
(contributing to an ${\rm N}^{v-2}$MHV amplitude),
the top-cut  is necessarily the
$v$-particle cut. This will always be $\eta$-independent
by construction.
Notice that this top cut will generically vanish if $v>4$.

{\bf 2.} All $p$-particle cuts which are generated
by the application of Feynman's Tree Theorem split
each one-loop MHV diagram into $p$ disconnected pieces
when $p > 1$. In all such cases
we see that amplitudes are produced
on all sides of the cut propagators when the sum over all
MHV diagrams is taken.

{\bf 3.} The case of a one-particle cut is  special since it
generates a connected tree diagram.
Similarly to the example discussed earlier,
one realises that by adding missing diagrams
the one-cut diagrams group into ${\rm N}^{v-1}$MHV amplitudes
with $n+2$ external legs (which are of course covariant).

{\bf 4.}
To see amplitudes appearing on all sides of the cuts,
one has  to sum over all one-loop MHV diagrams.

In this way one can show the covariance of  the result of a MHV diagram calculation 
for amplitudes with arbitrary helicities.

\subsection{Collinear   limits}

In this section we address the issue of reproducing the correct singularities
of the scattering amplitudes from MHV diagrams. 
We begin with collinear limits. At tree level, collinear limits were studied 
in \cite{csw1}, and shown to be in agreement with expectations from field theory. 
Consider now a one-loop scattering amplitude
$\mathcal{A}_n^{1-{\rm loop}}$. When the massless legs
$a$ and $b$ become collinear, the amplitude factorises
as \cite{bdk1,bdk2,Bern:1995ix,david}
\begin{eqnarray}
\label{coll}
&& \mathcal{A}_n^{1-{\rm loop}}
(1,\ldots, a^{\lambda_a}, b^{\lambda_b}, \ldots, n)
\;
{\buildrel a \parallel b\over
{\relbar\mskip-1mu\joinrel\longrightarrow}}
\\ \nonumber
&& \hspace{1 cm}\sum_{\sigma} \bigg[
 {\rm Split}^{\rm tree}_{- \sigma} ( a^{\lambda_a}, b^{\lambda_b})
\ \mathcal{A}_{n-1}^{1-{\rm loop}}
(1,\ldots, (a+ b)^{\sigma},  \ldots, n)
\nonumber \\
&& \hspace {1.4cm}
\ + \
{\rm Split}^{1-{\rm loop}}_{- \sigma} ( a^{\lambda_a}, b^{\lambda_b})
\ \mathcal{A}_{n-1}^{\rm tree}
(1,\ldots, (a+ b)^{\sigma},  \ldots, n)
\bigg]
\ .
\nonumber
\end{eqnarray}
${\rm Split}^{\rm tree}$ are the gluon tree-level
splitting functions, whose explicit forms can be found
e.g.~in  \cite{lance}. 
${\rm Split}^{1-{\rm loop}}$ is a
supersymmetric one-loop splitting function.
In \cite{ku} and \cite{vittorio}
explicit formulae for this one-loop splitting
function, valid to all orders
in the dimensional regularisation
parameter $\epsilon$, were found.
The result of \cite{vittorio} is:
\begin{equation}
\label{pippi}
{\rm Split}^{1-{\rm loop}}_{- \sigma} ( a^{\lambda_a}, b^{\lambda_b})
\ = \
{\rm Split}^{\rm tree}_{- \sigma} ( a^{\lambda_a}, b^{\lambda_b})
\ r(z)
\ ,
\end{equation}
where, to all orders in $\epsilon$, $r(z)=$
\begin{equation} 
\label{cl}
{c_\Gamma \over \epsilon^2} \Big( {-s_{ab} \over \mu^2} \Big)^{-\epsilon}
\left[  1 \, - \,
\mbox{}_{2}F_1 \left( 1, -\epsilon, 1- \epsilon, {z-1 \over z}\right)
  \, - \, \mbox{}_{2}F_1 \left(1, -\epsilon, 1- \epsilon, {z \over z-1}\right)
\right]
\end{equation}
and
\begin{equation}
c_\Gamma \ = \ {\Gamma (1 + \epsilon) \Gamma^2 ( 1 -  \epsilon) \over (4\pi)^{2- \epsilon}
\Gamma(1 - 2 \epsilon)}
\ .
\end{equation}
Here the parameter $z$ is introduced via the relations 
 $k_a := z k_P$, $k_b := (1-z) k_P$, where 
$k_P^2 \to 0$   in the collinear limit.
Notice that in \eqref{coll} we sum over the
two possible helicities $\sigma=\pm$.

In \cite{bst2} we were able to
reproduce \eqref{coll} from a calculation based on one-loop MHV diagrams; 
in particular we were able to re-derive the all-orders in $\epsilon$
expressions \eqref{pippi} and \eqref{cl}. 
Similarly to the tree-level, the different collinear limits at
one-loop arise from different MHV diagrams.

As an example, consider the  $++ \to +$ collinear limit. 
Consider first the diagrams where the two legs becoming collinear, $a$ and $b$, 
are either a proper subset of the legs attached to
a single MHV vertex,  or belong to a
four-point MHV vertex $\mathcal{A}_{4, {\rm MHV}} ( a, b, l_2,- l_1)$
but the loop legs $L_1$ and $L_2$ are connected to different
MHV vertices.%
\footnote{Thus, even if $L_1$ and $L_2$
become null and collinear, nothing special happens to the
sum of tree MHV diagrams on the right hand side of this
four-point MHV vertex,
precisely because $L_1$ and $L_2$ are not part of
the same MHV vertex.} 
In this case we call  $s_{ab}$  a {\it non-singular} channel.
Summing over all MHV diagrams
where $s_{ab}$ is a non-singular channel,
one immediately sees that
a contribution identical to the first term in
\eqref{coll} is generated. 

Next we consider  singular-channel diagrams 
(a prototypical one is shown in Figure 6),
i.e.~diagrams where  the legs $a$ and $b$  belong to a
four-point MHV vertex and the two remaining loop legs are
attached to the same MHV vertex. 
\begin{figure}[ht]
\label{Figure9}
\begin{center}
\scalebox{0.60}{\includegraphics{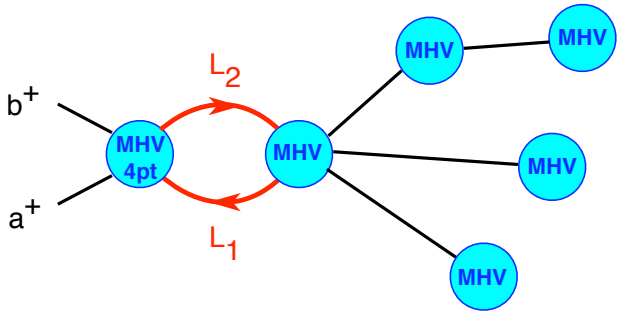}}
\end{center}
\caption{
A schematic example of a one-loop
MHV diagram contributing to a generic
non-MHV one-loop amplitude where
$s_{ab}$ is a singular channel. In the collinear
limit $a \parallel b$, diagrams of this type
generate the second term on the right hand side of  
\eqref{coll}.
}
\end{figure}
The diagram shown  in Figure 6 has been calculated in 
\cite{bst}, and in \cite{bst2} we showed that it precisely accounts 
for the second term in  \eqref{coll}. 

A similar analysis can be carried out for the other collinear limits, as well as for 
the soft limits. In all cases we found complete agreement with the known expressions 
for the all-orders in $\epsilon$ collinear and soft functions \cite{bst2}.

\section{Generalised Unitarity}

One-loop scattering amplitudes in
supersymmetric gauge theories are linear combinations of scalar
box, tensor triangle and tensor bubble integral functions with
coefficients that are rational functions of the momenta and spinor
variables. One wonders if these coefficients can be
determined directly without performing any loop integration. The
answer turns out to be yes and the main tool are unitarity
(2-particle) cuts and generalised (3-particle and 4-particle) cuts
\cite{Cutkosky:1960sp,bible}. In an $n-$particle cut of a loop
diagram, $n$ propagators are replaced by $\delta$-functions which
(partially) localise the loop integration and produce a sufficient
number of linear equations to fix all coefficients. The method
using 2-particle cuts was introduced \cite{bdk1,bdk2}, where it
was used to calculate the first infinite series of one-loop
amplitudes in SYM, and later generalised to include 3-particle
cuts in \cite{a,bdk-december}. Finally, in \cite{bcf-gen}
4-particle cuts were shown to be an efficient tool to find
coefficients of box functions. In particular they reduce the
calculation of all one-loop amplitudes in $\mathcal{N}=4$ SYM to a
simple algebraic exercise.

In non-supersymmetric theories loop amplitudes contain additional
rational terms, which do not have cuts in 4 dimensions. This
problem can be elegantly solved by working in $D=4-2
\epsilon$ dimensions, since then also rational terms develop cuts
and become amenable to unitarity techniques \cite{bm}. In
\cite{bmst} generalised unitarity techniques in $D$ dimensions
were shown to be a powerful tool to
calculate complete one-loop amplitudes 
in non-supersymmetric theories like QCD. The price to pay
is that we have to evaluate cuts in $D$ dimensions; in practice
this means that we have to make the internal particles massive
with uniform mass $\mu$ and integrate over $\mu$ \cite{bm}.

We now illustrate this technique for the simplest one-loop
amplitude in pure YM which has four positive helicity gluons
$\langle1^+2^+3^+4^+\rangle$. It turns out that in this case we
only have to consider the 4-particle cut depicted in Figure 7; no
other cuts are necessary.
\begin{figure}[h]
\begin{center}
\scalebox{0.60}{\includegraphics{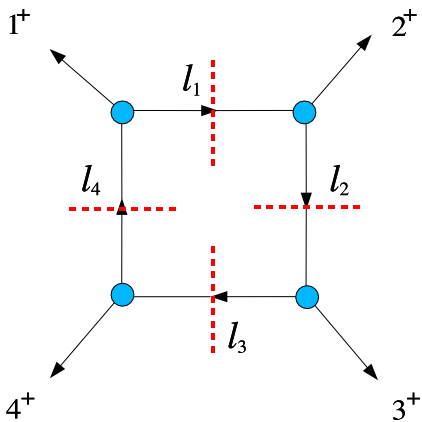}}
\caption{The 4-particle cut of the one-loop
$\langle1^+2^+3^+4^+\rangle$ amplitude in Yang-Mills}
\label{fig:largenenough}
\end{center}
\end{figure}
The 4-particle cut gives $\mu^{4} \frac{[12][34]}{\langle 12
\rangle \langle 34 \rangle}$ where $\mu$ is the mass of the
internal particle. This implies that the amplitude is proportional
to a scalar box integral with $\mu^{4}$ inserted in the integral
$I_{4}[\mu^{4}] = -\epsilon (1-\epsilon) I_{4}^{D=8-2\epsilon} =
-\frac{1}{6} +\mathcal{O}(\epsilon)$. Hence, the amplitude is 
purely rational, and we find
\begin{equation}
A_{4}(1^{+},2^{+},3^{+},4^{+})= -\frac{1}{6}
\frac{[12][34]}{\langle 12 \rangle \langle 34 \rangle} \, .
\end{equation}
In \cite{bmst} this method was applied successfully to the
remaining 4-point amplitudes $\langle1^+2^+3^+4^-\rangle$,
$\langle1^-2^-3^+4^+\rangle$ and $\langle1^-2^+3^-4^+\rangle$, and
to the particular 5-point amplitude
$\langle1^+2^+3^+4^+5^{+}\rangle$. Interestingly, it turned out
that for the amplitudes with one or more negative helicity gluons, only 
3-particle and 4-particle cuts were needed in order to determine the amplitude. 
This method can be applied directly to more
general amplitudes and the 5- and 6-point amplitudes 
are currently under investigation.

\section*{Acknowledgments}
We would like to thank Bill Spence, James Bedford and Simon McNamara for 
very pleasant collaborations, and 
Adi Armoni, Zvi Bern, Emil Bjerrum-Bohr, Freddy Cachazo, 
Luigi Cantini, Lance Dixon, Dave Dunbar, Michael Green, 
Harald Ita, Valya Khoze, David Kosower, Paul Mansfield, 
Marco Matone, Christian R\"{o}melsberger  and Sanjaye Ramgoolam  
for discussions.
The work of AB is partially
supported by a PPARC Special Project Grant.
The research of GT is supported by an EPSRC Advanced  Fellowship.

\end{document}